\input{aipcheck}

\documentclass[
     ,final            
  ]
  {aipproc}

\layoutstyle{6x9}

\begin{document}
\title{3-D Simulations of MHD Jets \\ 
- The Stability Problem}

\author{Masanori Nakamura and David L. Meier}
{address={Jet Propulsion Laboratory, 
California Institute of Technology, Pasadena, CA 91109, USA}}

\begin{abstract}
Non-relativistic three-dimensional magnetohydrodynamic simulations of
Poynting-flux-dominated (PFD) jets are presented. 
Our study focuses on the propagation of strongly magnetized hypersonic 
but sub-Alfv\'enic flow ($C_{\rm s}^2 << V_{\rm jet}^2 < V_{\rm A}^2$) 
and the development of a current-driven (CD) kink instability.
This instability may be responsible for the "wiggled" structures 
seen in VLBI-scale AGN jets.  
In the present paper we investigate the nonlinear behavior of PFD jets 
in a variety of external ambient magnetized gas distributions, 
including those with density, pressure, and temperature gradients. 
Our numerical results show that PFD jets can develop kink distortions 
in the trans-Alfv\'enic flow case, even when the flow itself is still 
strongly magnetically dominated.  
In the nonlinear development of the instability, a non-axisymmetric mode 
grows on time scales of order the Alfv\'en crossing time (in the jet frame) and 
proceeds to disrupt the kinematic and magnetic structure of the jet. 
Because of a large scale poloidal magnetic field in the ambient medium, 
the growth of surface modes ({\it i.e.}, MHD Kelvin-Helmholtz 
instabilities) is suppressed. 
The CD kink mode ($m = 1$) grows faster than 
the other higher order modes ($m > 1$), driven in large part by 
the radial component of the Lorentz force. 
\end{abstract}

\maketitle


\section{Introduction}
Magnetohydrodynamic (MHD) mechanisms are commonly considered
to be the likely model for accelerating 
the wind/outflow from Young Stellar Objects (YSOs),
X-ray binaries (XRBs), Active Galactic Nuclei (AGN), Microquasars, and
Quasars (QSOs) \citep[see, e.g.,][and references therein]{ME01}.
The toroidal (azimuthal) component of the magnetic field, generated 
by the twisting of the magnetic field threading an accretion disk, plays 
an important role in magnetic acceleration \citep{US85}.
High resolution observations of AGN show morphological
structures, such as ``wiggles (kinks)'' or ``bends'', not only on 
kpc scales, but also pc scales and smaller \citep{HU92}.
Such helical distortions might be caused either by plasma instabilities 
or by precession of the jet ejection axis due to the gravitational interaction 
of binary Black Holes (BBHs) \citep{BE80}, BH/disk, or galaxies. 

The magnetically driven jet model may be supported by recent observations.
The rotation-measure (RM) distribution for the parsec-scale 3C 273 jet 
has a systematic gradient across the jet, and the projected magnetic 
field vector is systematically tilted from the jet central axis \citep{AS02}. 
Similar RM gradients also occur in several BL Lac objects \citep{GM03}.
These results 
indicate the presence of a toroidal component of the field inside jet. 

Based on both theoretical aspects and observational results, it is 
natural to investigate plasma instabilities that might occur in these 
observed morphological structures.
In the MHD model the toroidal field component provides the collimation of 
the jet by magnetic tension or ``hoop'' stress.
However, it is well-known 
that such a cylindrical plasma column with a helical magnetic configuration 
is subject to MHD instabilities.
These are usually divided into pressure-driven instabilities (PDIs) and 
current-driven instabilities (CDIs) \citep[see, e.g.,][]{BA80}.
CDIs arise because the toroidal field component of a 
magnetically driven jet is equivalent to a poloidal (axial) electric current,
and such ``current-carrying'' jets are susceptible to MHD instabilities.
Kelvin-Helmholtz instabilities (KHIs), which are driven by velocity gradients,
must be also taken into account \citep[see, e.g.,][]{CH61}. 
Because these above instabilities would occur together in real astrophysical 
situations, it might difficult to separate their individual effects.

In the recent past, much less theoretical attention has been paid by the 
astrophysical community to PDIs \citep{BE98, KE20} and CDIs \citep{EI93, SP97, BE98} 
than to KHIs.
KHIs have been considered for 
the past few decades by many theoretical or numerical investigators 
\citep[for reviews, see][and references therein]{BI91, FE98}.
In general, sub-Alfv\'enic jets ($V_{\rm j} < V_{\rm A}$) are stable 
against KHIs, while, super-Alfv\'enic, but trans-fast-magnetosonic jets 
($V_{\rm A} < V_{\rm j} \sim V_{\rm FM}$) are KH unstable. 
For super-fast-magnetosonic jets, if they have a considerably large 
fast-magnetosonic Mach number $M_{\rm FM} (\equiv V_{\rm j}/V_{\rm
FM})$, then the jets gradually become stable again.
However, there are several effects that can stabilize jets beyond the 
Alfv\'en surface: (1) toroidal (rotational) velocity \citep{BO96}
and toroidal magnetic field \citep{LPMB89, AC92},
(2) considerable mass entrainment which increases 
$\eta$ ($\equiv \rho_{\rm j}/\rho_{\rm ext.}$) 
\citep{TO92, RH00}, and (3) external magnetic fields \citep{TO92}
or magnetized winds \citep{HR02, HH03}.
Also, as found by \citep{OCP03}, jets can maintain their stability 
by keeping the average Aflv\'en Mach number 
($M_{\rm A} \equiv V_{\rm j}/V_{\rm A}$) within the jet to order 
unity by concentrating poloidal magnetic flux near the jet axis, 
giving it a reasonably stiff ``backbone''. 

\citet{ALB00} investigated cold super-fast-magnetosonic 
force-free jets ($V_{\rm FM} < V_{\rm j}$). 
In the linear phase, the fastest growing CDI ($m=1$) is nearly 
independent of the radial profile of the pitch.
\citet{LBA00} then performed numerical simulations based on 
their linear analysis. 
In the early non-linear phase, nothing in 
their numerical results indicate a possible disruption of jets 
due to CDIs ($m=0 ~ {\rm and} ~ 1$). 
To our knowledge, only two numerical studies of pure 
CDIs for astrophysical jets have been performed. \citet{TO93} 
applied their results to YSO jets, and \citet{NUS01} 
applied them to a large scale (kpc scale) wiggled structure of AGN jets.

In the present paper, we report 3-D non-relativistic MHD
simulations of PFD jets and investigate CDIs in decreasing
(density, pressure, and temperature gradients) magnetized atmospheres.
PFD jets should be applicable to small scale AGN jets, such as those on parsec 
scales and smaller.
We believe that the helical morphological features 
(wiggles and bends) indicate the presence of CDIs in highly magnetically 
dominated jets.

\section{Numerical Models}
The numerical approach in the present paper assumes non-relativistic 
compressible ideal MHD, with gravity neglected.
We simulate the non-linear system of time-dependent MHD equations 
in a 3-D Cartesian coordinate system ($x,y,z$).
All physical quantities are normalized with a unit scale length $L_{0}$, 
typical density $\rho_{0}$, typical velocity $V_{0}$, and combinations thereof.
$\rho_{0}$ and $V_{\rm A 0}$ are the initial value of the density 
and Alfv\'en velocity at the origin of the computational domain. 
$L_{0}=2 R_{\rm j0}$ is the initial jet diameter at $z=0$. 

The initial distribution of the external ambient inter-galactic medium (IGM) 
consists of a stratified and magnetized atmosphere with possible gradients 
in density, pressure, and temperature.
We adopt an initial current- (and therefore force-) free magnetic configuration, 
$\mbox{\boldmath$J$} 
(\equiv \nabla \times \mbox{\boldmath$B$})=0$, 
where $\mbox{\boldmath$B$}$ is the magnetic field and
$\mbox{\boldmath$J$}$ the corresponding current density.
Under the assumptions of flux and mass conservation, the density 
will be proportional to the local magnetic field as $\rho \propto
|\mbox{\boldmath$B$}|^{\alpha}$, where $\alpha$ is a free parameter.
For $\alpha=2$, the Alfv\'en speed is constant throughout the computational 
domain, while for $\alpha < 2$, the Alfv\'en speed decreases with 
distance from origin and for $\alpha > 2$ it increases. 
The gas pressure $P$ is assumed to be polytropic,
$P \propto \rho^{\Gamma}$, where $\Gamma=5/3$ is the polytropic index. 
The atmosphere is artificially bound to prevent it from expanding
under its own pressure gradient by introducing a
``pseudo-gravitational'' potential designed to hold on to the atmosphere
without significantly impeding the advancing jet \citep{CHC97}.
In the lower ``boundary zone'' ($z_{\rm min}<z<0$) the boundary velocity 
is specified for all time as 
$\mbox{\boldmath$v$}_{\rm j}=v_{\phi} (r,z) \hat{\phi}+v_{z} (r,z)
\hat{z}$, where $r=(x^2+y^2)^{1/2}$.
This represents a continuous cylindrical MHD inflow,
powered by non-linear torsional Aflv\'en waves (TAWs) into the 
``evolved'' region ($z \geq 0$) of the computational domain.
The total computational domain is taken to be $|x| \leq x_{\rm max}$, 
$|y| \leq y_{\rm max}$, and $z_{\rm min} \leq z \leq z_{\rm max}$, 
where $x_{\rm max}, \ y_{\rm max} \simeq 16 \ (32R_{\rm j0})$, $z_{\rm
min} \simeq -1 \ (2R_{\rm j0})$, and $z_{\rm max} \simeq 20 \ (40R_{\rm j0})$.
The numbers of grid points in the simulations reported here are 
$N_{x}\times N_{y}\times N_{z}=261 \times 261 \times729$, where the grid
points are distributed non-uniformly in the $x$, $y$, and $z$ directions.
High-resolution 3-D computations were performed on a FUJITSU VPP 5000/32R
(9.6 GFLOP/s peak speed per processor) and required about 6 hours on the
32-processor machine.

\section{Numerical Results}
We concentrate on the solution in the region, $-2.25 \leq x,~ y \leq 2.25$, 
and $0.0 \leq z \leq 18.0$ and discuss the dynamical behavior of a typical case: 
the growth of CDIs in a PFD jet, with parameter $\alpha = 1$.
$B_{z}$ and $\rho$ decrease gradually along $z$-axis and tend 
to $\sim z^{-2}$ for $z > 1$.
We set the plasma beta ($\beta_{0} \equiv 2 C_{\rm s0}^{2}/\gamma V_{\rm A0}^2 = 10^{-2}$) 
at the origin, and the ratio $V_{\rm A}/C_{\rm s}$ increases gradually along the $z$-axis.
Throughout the time evolution, a quasi-stationary PFD flow 
($F_{E \times B}/F_{\rm tot.} \sim 0.9$, where 
$F_{E \times B}$ Poynting flux and 
$F_{\rm tot.}$ total energy flux) is injected into 
the ``evolved'' region from the lower ``boundary'' zone.
\begin{figure}[htbp]
\includegraphics[scale=0.75,angle=0]{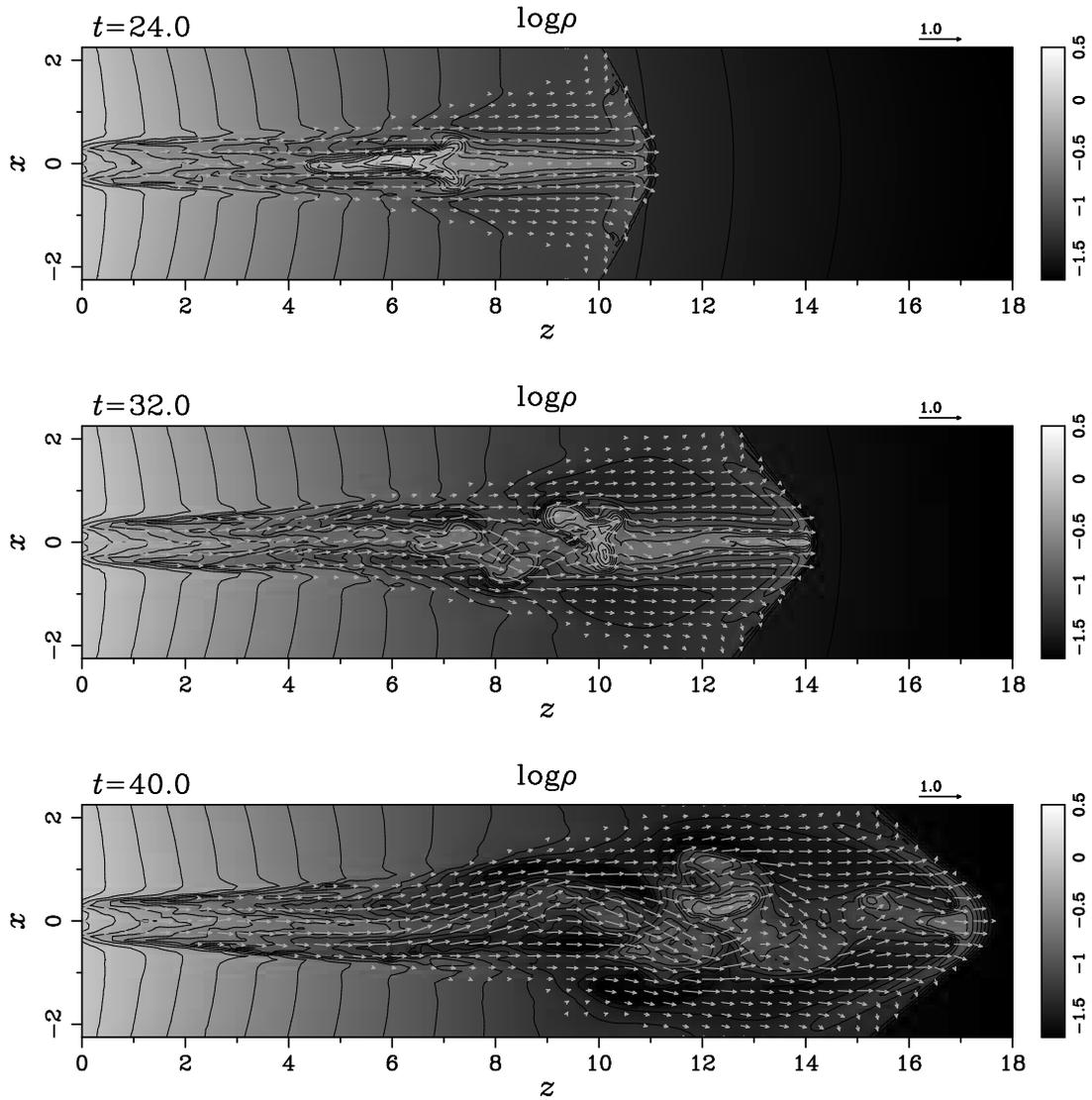}
\caption{Time evolution of a typical PFD jet. Gray-scale images of the density distribution 
(logarithmic scale) are shown with the poloidal velocity field in the $x-z$ plane at
$t=24.0$ ({\it top}), $t=32.0$ ({\it middle}), and $t=40.0$ ({\it bottom}). 
\label{fig1}}
\end{figure}
Fig. 1 shows the time evolution of the density and the velocity field in the $x-z$ plane
for the typical PFD jet.
In the early stage ($t=24.0$), a strongly magnetized helical jet powered by
TAWs advances into the decreasing poloidally magnetized IGM, {\em remaining in radial
force-free equilibrium} ($ \mbox{\boldmath$J$} \times \mbox{\boldmath$B$} \simeq 0$).
This is precisely the transverse structure predicted analytically by \cite{LPMB89}. 
The front of the TAW (a fast-mode MHD wave) is decelerated due to
gradually decreasing $V_{\rm A}$; the accumulation of $B_{\phi}$
occurs behind this wave front.
The toroidal magnetic pressure gradient force $d (B_{\phi}^{2}/8 \pi) /dz$
becomes large and works effectively at this front, strongly compressing 
the external medium along this propagating TAW front.
At later stages the front becomes super-fast magnetosonic and a bow
shock (MHD fast-mode shock) is formed.
The accumulation of $B_{\phi}$, due to the gradually decreasing $V_{\rm A}$,
causes a concentration of axial current $J_{z}$ near central ($z$) axis, 
and the distribution of magnetic pitch ($|B_{\phi}/r B_{z}|$) become large. 
The Lorentz force breaks the quasi-equilibrium balance in the radial direction,
and the axisymmetric PFD jet is disrupted by a CDI kink mode ($m = 1$) ($t=32.0$).
At late times ($t=40.0$, see Fig. 1 {\it bottom} and Fig. 2 {\it right})
a large-scale wiggled structure appears, while the MHD jet is still 
sub-Alfv\'enic $V_{\rm jet} < V_{\rm A}$ (except at the front itself;
see Fig. 2 {\it left}).
This PFD jet will continue to be subject to the CDI, even thought it is
magnetically dominated throughout the computational time.
\begin{figure}[htbp]
\includegraphics[scale=0.5,angle=-90]{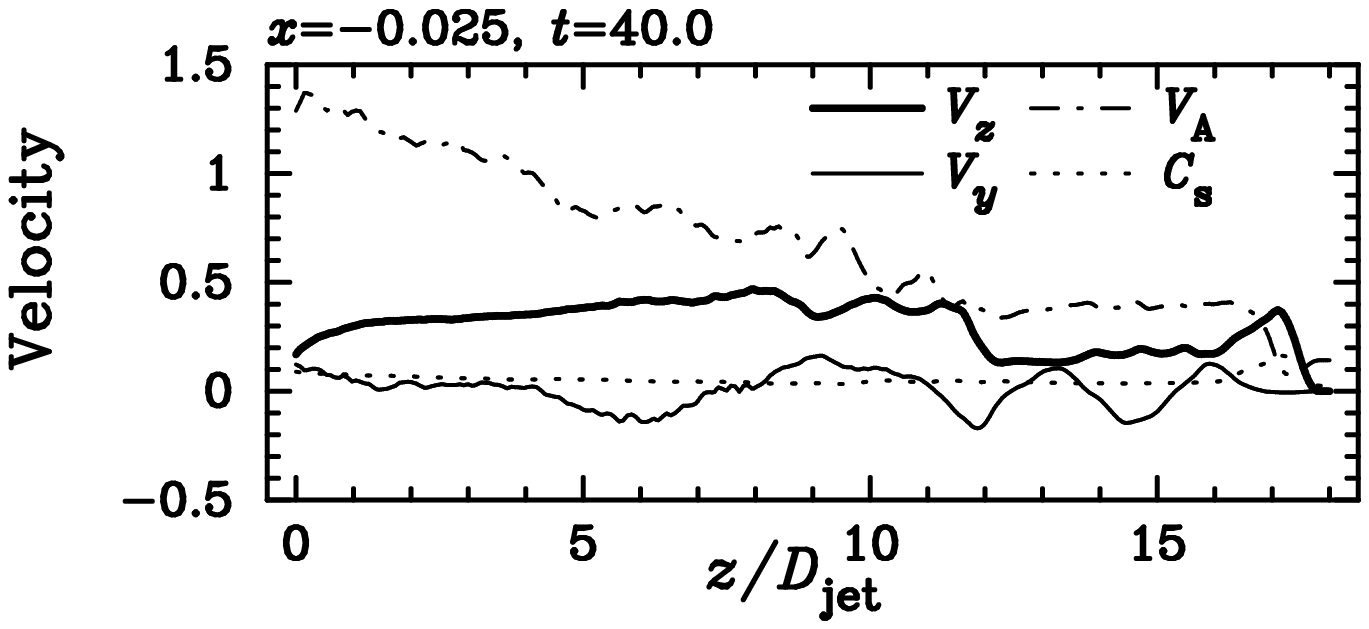} ~ ~ ~
\includegraphics[scale=0.5,angle=-90]{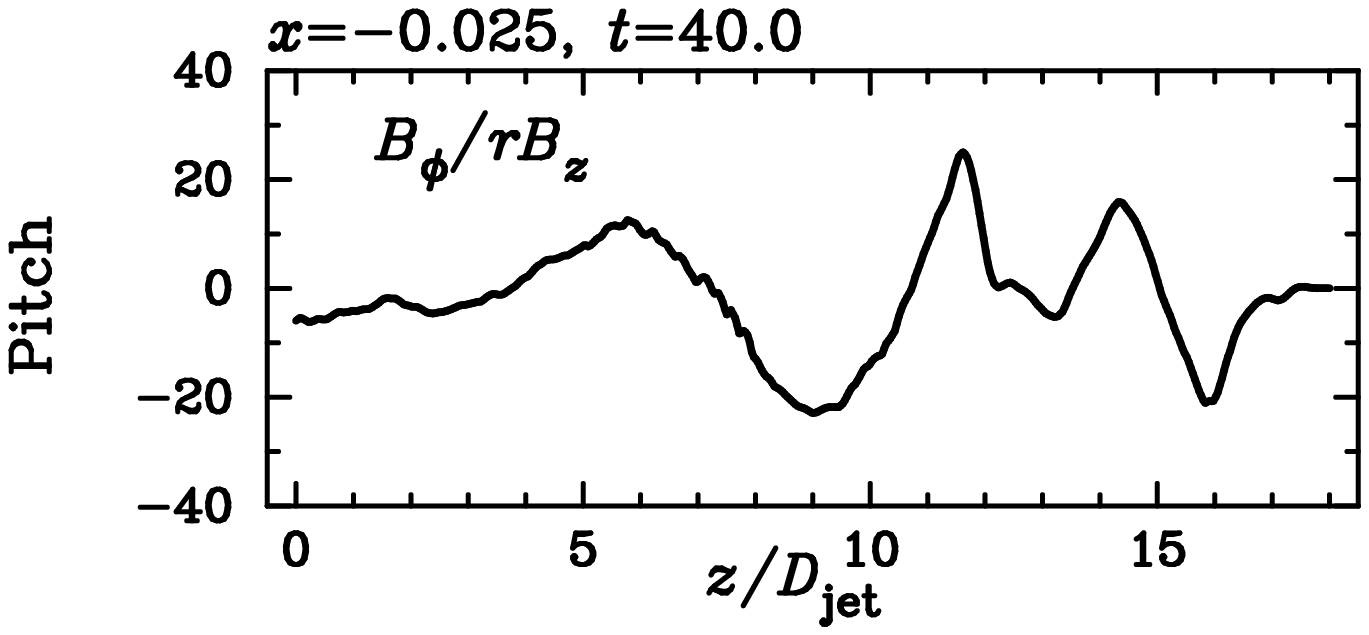}
\caption{Distribution of velocities ({\it left}) and magnetic pitch ({\it right}) 
parallel to the $z$-axis  \label{fig2}}
\end{figure}
Fig. 3 shows the closed circulating current system, consisting of 
a current that is co-moving with the PFD jet (close to central axis) and 
a return current that flows outside the magnetized wind ({\it cocoon}).
The external magnetized wind reduces the velocity shear between the jet
itself and external medium, thereby suppressing KHIs even if the MHD jets
become super-Alfv\'enic ({\it i.e.}, a kinetic energy flux-dominated [KFD]). 
\begin{figure}[htbp]
\includegraphics[scale=0.6,angle=-90]{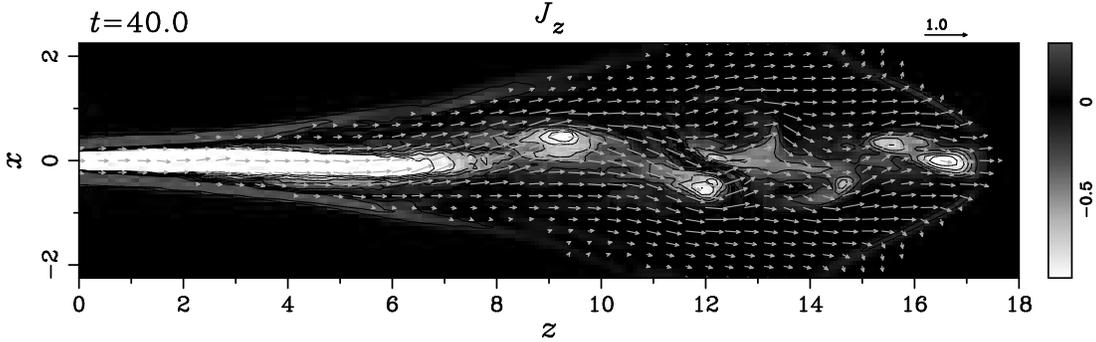}
\caption{Gray-scale image of the axial current density distribution in $x-z$
plane. \label{fig3}}
\end{figure}
\section{Discussion}
There now is observational evidence that, in at least some extragalactic 
systems, the AGN jets are slowly collimated over a length scale of pc ($>>$
BH + disk system) \citep[e.g., M87,][]{BJL02}.
If the MHD process is indeed the acceleration mechanism, 
the Alfv\'en surface will be at a distance of order the disk's outer radius
$R_{\rm disk}$.
The observations can set an upper limit on this of 
a few tens of pc (the radius of the ionized disk \citep{FO94}).
From theoretical considerations, the optically-thick disk is expected to
be at about $R_{\rm disk} \sim 20$ pc \citep{CF90}.
So, an understanding of sub-Alfv\'enic to trans-Aflv\'enic jet flow may be
needed in order to understand jet dynamics and morphology on sub-pc to pc
scales.

\section{Conclusions}
We have investigated the non-linear behavior of
PFD jets powered by TAWs in the extended stratified atmospheres, and 
find the disruption of PFD jets due to a CDI kink
mode, leading to the formation of wiggles.
Even if jets become super-Alfv\'enic during their propagation, 
CDI disruptions can still occur. 
Growing modes of MHD KHIs for super-Alfv\'enic flows are not seen 
in our parametric survey, because of external force-free magnetized helical
winds surrounding the well-collimated jet, suppressing the growth of KHIs.
CDIs might be a possible origin of the wiggled morphology
of AGN jets are slowly accelerated and collimated on pc scales. 
 

\begin{theacknowledgments}
We thank P. E. Hardee for useful comments and suggestions.
M.N. is supported by a National Research Council Resident Research
Associateship, sponsored by the National Aeronautics and Space Administration.
This research was performed at the Jet Propulsion Laboratory,
California Institute of Technology, under contract to NASA.
Numerical computations were performed on the Fujitsu VPP5000/32R 
at the Astronomical Data Analysis Center of the National Astronomical
Observatory, Japan.
Data analyses and visualizations were carried out 
at the Supercomputing and Visualization Facility at JPL.
\end{theacknowledgments}


\bibliographystyle{aipproc}   



\end{document}